\documentclass[reprint,aps,amsmath,amssymb,pre,twocolumn]{revtex4-2}
\usepackage{graphicx}
\usepackage{dcolumn}
\usepackage{bm}
\usepackage{nameref}
\usepackage{hyperref}
\usepackage{multirow}
\usepackage{xcolor}

\begin{document}
\title{Theory of Semi-discontinuous DNA Replication}
\author{Janani G}
\author{Deepak Bhat}%
 \email{deepak.bhat@vit.ac.in}
\affiliation{Department of Physics, School of Advanced Sciences, Vellore Institute of Technology, India}
\date{\today}

\begin{abstract} The replisome, a protein complex that drives DNA replication, comprises multiple DNA polymerases. During replication, one polymerase synthesizes the leading strand continuously, while another synthesizes the lagging strand discontinuously, producing short transient segments known as Okazaki fragments and gaps between them. After the synthesis of each Okazaki fragment, the lagging-strand polymerase dissociates either spontaneously or upon collision with the preceding fragment; however, the relative contributions of these mechanisms remain unclear. Here, we develop a biophysical model of semi-discontinuous replication by incorporating the stochastic dynamics of the lagging-strand polymerase. By computing the Okazaki fragment and gap-size distributions, we show the significance of polymerase dissociation in shaping these statistics. Applying the model to the T4 bacteriophage and \textit{B. subtilis} replisomes, we find that collisions with preceding Okazaki fragments predominantly trigger polymerase dissociation. \end{abstract}


\maketitle

\section{Introduction}
Biological cells have a remarkable ability to replicate their genome. Genome replication is facilitated by the replisome, a protein complex containing more than ten proteins. The primary proteins of a replisome in all forms of life are helicase, primase, clamp loader, and multiple DNA polymerases. These proteins work in a coordinated fashion, form a Y-shaped fork by separating the two strands of a DNA template, and synthesize the complementary sequence on both strands of the DNA\cite{alberts1987prokaryotic,Johnson2005, Alberts2014,benkovic2001replisome,duderstadt2014replication,  Burgers2017}.

 Replisomes copy nearly 50\% of the genome discontinuously in every biological cell. The antiparallel orientation of the DNA strands and the inability of polymerase to copy in 5' to 3'  direction necessitate this semi-discontinuous replication \cite{benkovic2001replisome}. During replication, a polymerase continuously copies the leading strand (3' to 5') of DNA in the direction of the replisome\cite{Alberts2014,Yao2009,Yuan2013}. However, the lagging strand synthesis occurs opposite to the direction of the replisome, and the process is more complex: a primase synthesizes a primer in the vicinity of the replisome, a clamp loader loads a clamp to the primer, enabling a polymerase to bind to the primer and begin replication. Eventually, the lagging-strand polymerase dissociates from the strand after synthesizing a short fragment and becomes available for rebinding to synthesize the next fragment \cite{mueser2010structural}. The short fragments synthesized by the lagging-strand polymerase are known as the Okazaki fragments (OFs) \cite{okazaki1968mechanism}, and single-stranded regions between successive OFs are called gaps. The OFs and gaps are later processed by dedicated gap-filling proteins including specialized polymerases and joined together by DNA ligase \cite{williams2021high} and improper processing of OFs is linked to human pathologies \cite{zheng2011,sun2023okazaki}.

The sizes of OFs have been measured by varying the concentrations of polymerase,  primase, and clamp loader \cite{chastain2000analysis,chen2013insights,spiering2017rna,nelson2008rna}.   The OFs are typically of $1-3\, {\rm kbp}$ in prokaryotes and $200-300\, {\rm bp}$ in eukaryotes \cite{balakrishnan2010reconstitution}.

While semi-discontinuous replication is known to be a reason for differences in replication errors between the two DNA strands \cite{Katarzyna2018, reijns2015}, concomitant error removal facilitated by frequent dissociation of the lagging-strand polymerase is thought to be the primary cause for the differences \cite{Katarzyna2018}. Experimental studies have identified steric collision with the preceding OF \cite{hacker1994rapid} and pre-collision spontaneous dissociation \cite{yang2006control,Fay1982,tanner2008single} as potential mechanisms of lagging-strand polymerase dissociation; however, how often each mechanism operates during replication, relative to the other, remains unclear. 

OF sizes, typically measured using gel electrophoresis or electron microscopy \cite{chastain2000analysis,chen2013insights,yang2006control}, are sensitive to lagging strand polymerase dissociation and can thus be leveraged to identify the dissociation mechanism. However, this inference requires a mathematical framework that bridges the microscopic dynamics of the lagging-strand polymerase with experimentally measured OF sizes. Most mathematical models of DNA replication have focused on replication initiation \cite{Retkute2011, Retkute2012, Karschau2012, Kelly2019, Moura2024}, exonuclease activity \cite{Sharma2013, Sahoo2021}, DNA damage \cite{Gauthier2010}, and speed variations along the genome \cite{Bhat2022}. Semi-discontinuous replication is ubiquitous across all forms of life and plays a crucial role in shaping the mutational landscape of both strands of DNA. Nevertheless, a general theory of semi-discontinuous replication remains elusive.

\begin{figure}
    \centering
    \includegraphics[width=0.95\linewidth]{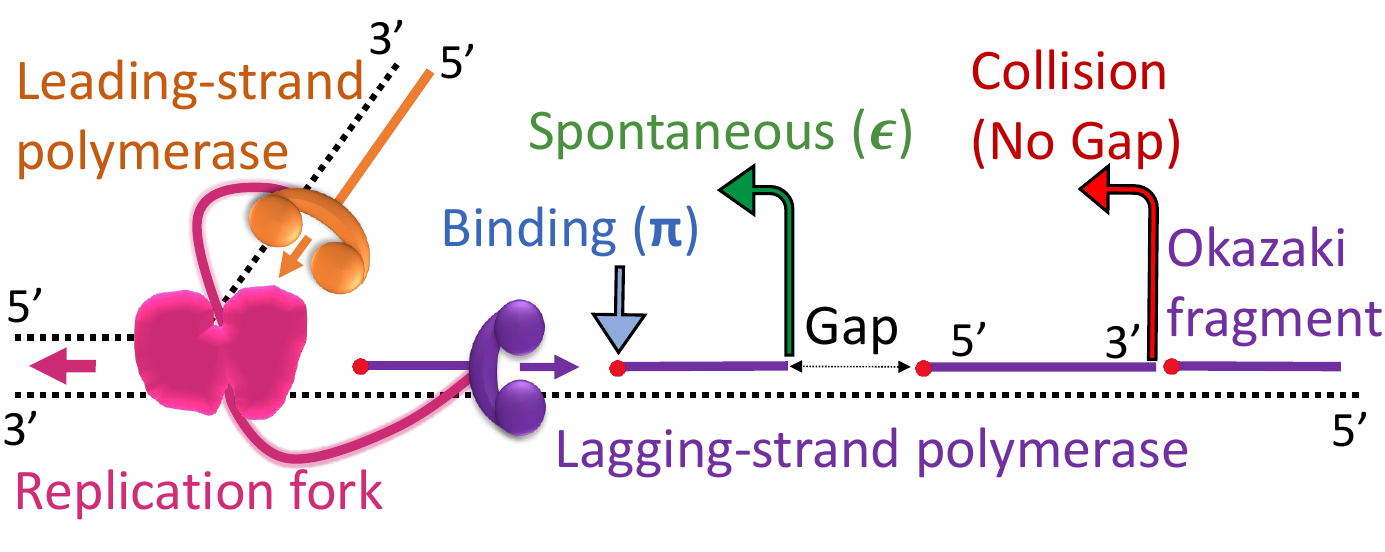}
    \caption{ A schematic view of our model for replisome carrying out semi-discontinuous replication.}
   \label{fig: Model picture}
\end{figure}

In this work, we propose a model for semi-discontinuous replication by considering the stochastic growth--reset dynamics of lagging-strand synthesis carried out by a DNA polymerase. By computing the size distribution of OFs and the fraction of collision-induced polymerase dissociation, we show that the distribution transitions from a monotonic form in low-collision regimes to a non-monotonic form in high-collision regimes. Analysis of the previously reported size distribution of OFs for T4 bacteriophage and \emph{B. subtilis}, suggests that steric collisions primarily regulate the length scales of OFs and gaps during DNA replication.

\section{Model}
In our model, we consider a replisome, consisting of two independent polymerases, that moves one-dimensionally along the negative x-axis at a uniform speed $v$ (Fig.~\ref{fig: Model picture}). A polymerase continuously copies the leading strand at the same speed and direction as the replisome; therefore, we treat their positions as identical. In contrast, another polymerase synthesizes OFs through stochastic growth and resetting dynamics on the lagging strand; in an unbound state, the polymerase binds the lagging strand with a rate $\pi$ in the vicinity of the replisome, replicates with the speed ($v$) in the direction opposite to the replisome, dissociates from the lagging strand after synthesizing an OF, and becomes available for rebinding. \\
We distinguish two types of polymerase dissociations from the strand. While synthesizing an OF, the polymerase can collide with a preceding OF and dissociate instantly. It can also dissociate \emph{spontaneously} with rate $\epsilon$ due to thermal fluctuations or other molecular interactions. Spontaneous dissociation leaves a gap on the lagging strand, while the collision does not, see Fig.~\ref{fig: Model picture}.
 
It is convenient to introduce rescaled parameters: $p=\pi/\epsilon$- the ratio of the binding rate to the spontaneous unbinding rate, $q=\epsilon/v$ -  (rescaled) spontaneous unbinding rate, and $r=\pi/v$ - (rescaled) binding rate. We express our results in terms of these parameters.

We characterize semi-discontinuous replication using three key quantities: (a) OF size distribution- $Q(z)$,  defined such that $Q(z)dz$ is the probability for an OF having size between $z$ and $z+dz$. (b) Gap-size distribution- $R(g)$, defined such that $R(g)dg$ is the probability for a gap having size between $g$  and $g+dg$. (c) The fraction of collision, $f_c$, is the fraction of dissociations of the lagging-strand polymerase caused by collisions with the preceding OF, rather than spontaneous unbinding.

\section{Main results}
\begin{figure}
	\centering
    	\includegraphics[angle=270,scale=0.18]{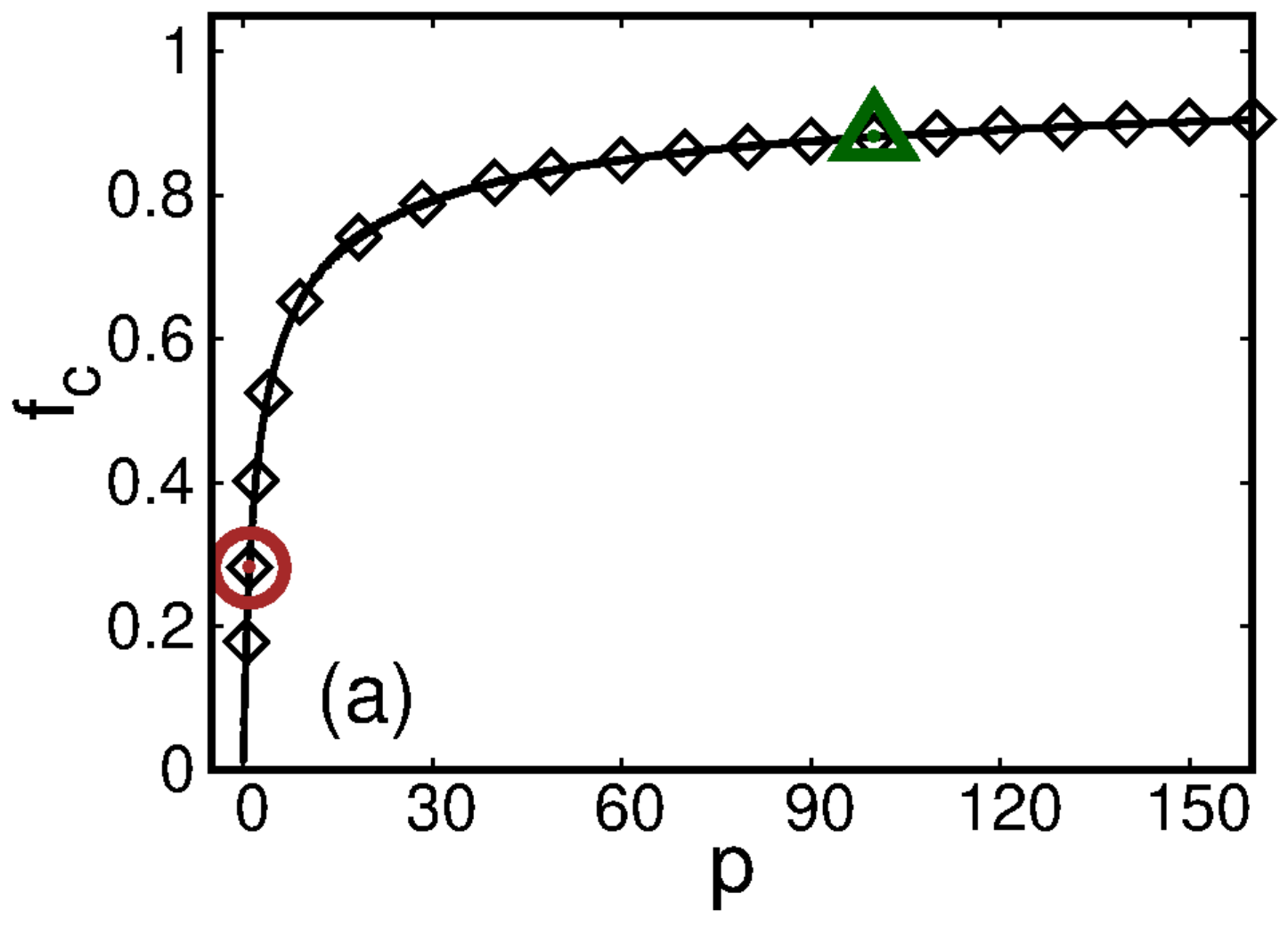}
	\includegraphics[angle=270,scale=0.167]{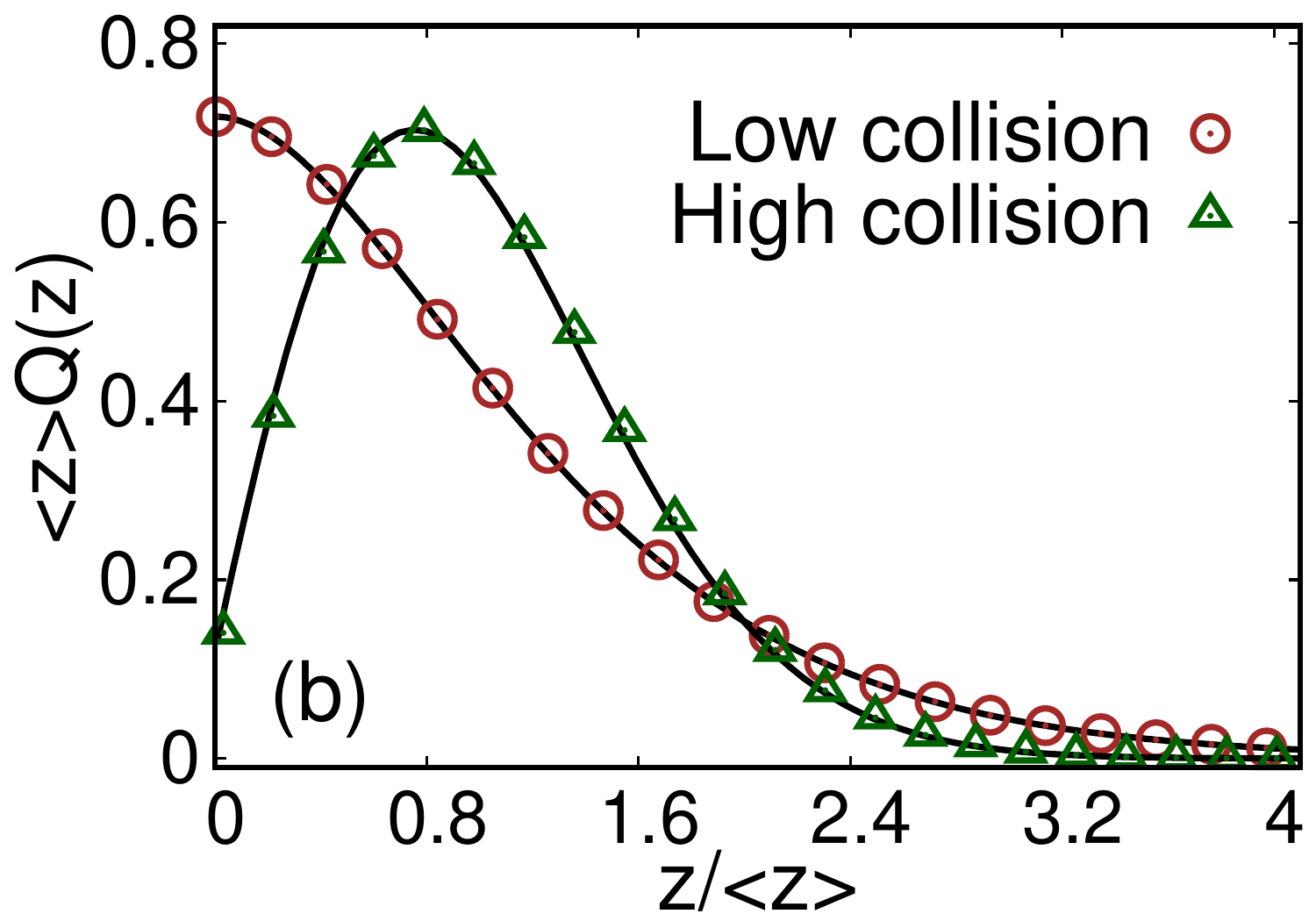}
    \includegraphics[angle=270,scale=0.167]{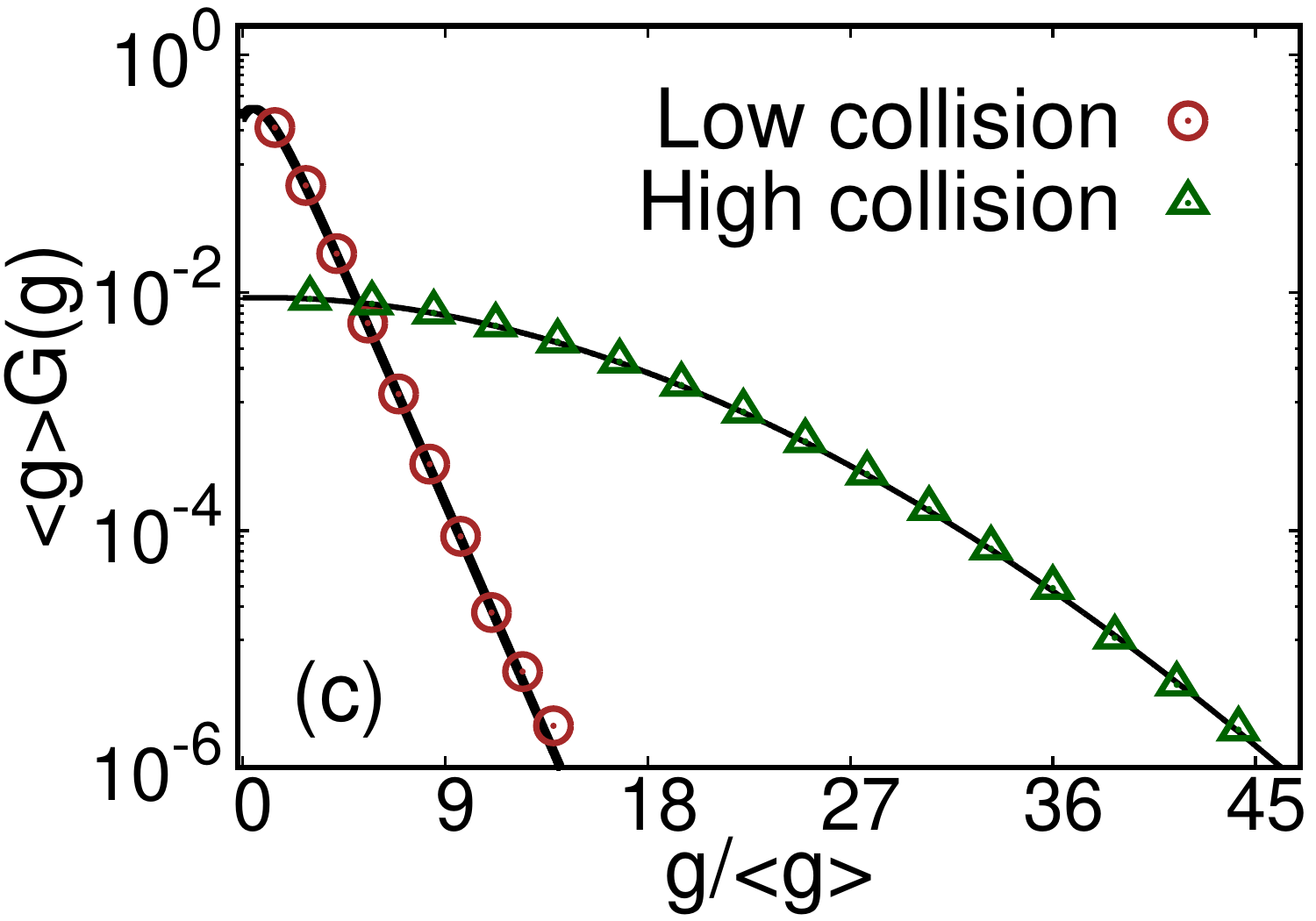}
	\caption{Effect of collision-induced polymerase dissociation on the OF and gap sizes: (a) The fraction of collision (Eq.~\ref{eq:eq2}) increases with $p$. (b) The OF size distribution  (Eq.~\ref{eq:eq1}) is monotonic for low collisions ($p=1$) and non-monotonic at high collisions ($p=100$). (c) The gap-size distribution is exponential at low collisions and Gaussian-like at high collisions. Symbols denote simulation results. In all simulations, we fixed $\epsilon=0.05 \,{\rm sec^{-1}}$  and $v=400 \,{\rm bp \cdot sec^{-1}}$.  In (a), $\pi$ is varied from $0.005\, {\rm sec^{-1}}$ to $8.5\, {\rm sec^{-1}}$. In (b) and (c), we use $\pi=0.05 \, {\rm sec^{-1}}$ and $\pi=5 \, {\rm sec^{-1}}$ corresponding to $p=1$ and $p=100$, respectively. In (a), $f_c$ is corresponding to the same parameter sets as in (b) and (c) is highlighted using enlarged symbols.} 
    \label{fig:overall}
\end{figure}

We now outline our main results. The size distribution of the OFs is given by
\begin{align}
Q(z) = A(z)~e^{-\int_0^{z}A(z')dz'}
\label{eq:eq1}
\end{align}
where $A(z)=(q+r)-r e^{-q z}$. The distribution is monotonic for $r<q$ and non-monotonic for $r>q$, with a transition at $r=q$, see Fig.~\ref{fig:overall}(b). The non-monotonicity arises from the collision-induced polymerase dissociation by the preceding OFs. To quantify this, we compute the fraction of collisions,
\begin{align}
f_c \;=\; 
\, 1 - \frac{e^p}{p^{1+p}}\,
\gamma\left(1+p,\,p\right) \,,
\label{eq:eq2}
\end{align}
where $\gamma(a,b) = \int^{b}_{0} t^{a-1}e^{-t}dt$ is the incomplete Gamma function. Invoking the approximation, $\gamma(1+p,p) \approx p^{1+p}e^{-p}/(1+p)$, we obtain an intelligible form, $f_c\approx p/(1+p)$. Thus, $ f_c\to 0$ as $p\to 0$, indicating that nearly all dissociations are spontaneous, while $f_c\to 1$ as $p\to \infty$, corresponding to a collision-dominated regime (see Fig.~\ref{fig:overall}(a)).

To analyze the role of collision-induced dissociation in shaping the OF size distribution, consider the limit $p \to 0$, where the binding rate is significantly lower than the spontaneous unbinding rate. In this limit, the initiation sites of successive OFs are sparsely distributed, so the polymerase typically dissociates spontaneously, yielding $f_c \approx 0$. Consequently, the OF size distribution follows a monotonic exponential decay, $Q(z) \sim q e^{-q z}$, see Fig.~\ref{fig:overall}(b). As the binding rate increases, the inter-fragment distance decreases, driving the fraction of collisions ($f_c$) to $\sim 28\%$ for $p = 1$, to exceed $50\%$ for $p > 3.5$, and to approach $100\%$ asymptotically ($p \to \infty$). In this collision-dominated regime, the spontaneous dissociation is rare, and the processivity is limited by the downstream OF.  This constraint suppresses the very long fragments and shifts the OF population toward intermediate lengths, resulting in the characteristic non-monotonic profile of $Q(z)$ observed in Fig.~\ref{fig:overall}(b).

A direct influence of collisions may be seen on the mean OF size, defined as $\langle z \rangle = \int_0^{\infty} z Q(z)dz$. Computing the integral from Eq.~\ref{eq:eq1} and expressing it in terms of $f_c$ given in Eq.~\ref{eq:eq2}, we  obtain
\begin{align}
 \langle z \rangle = \frac{1-f_c}{q}\,.
 \label{eq:eq3}
\end{align}
For a fixed spontaneous unbinding rate, increasing the fraction of collisions decreases the mean fragment size. 

Complementary to the fragment sizes, the lagging-strand replication is further characterized by the gap-size distribution,
\begin{align}
    R(g) = \delta(g) f_c + G(g)\,.
      \label{eq:eq4}
\end{align}
where the delta-function term represents zero-sized gaps generated by collisions, and hence the weight $f_c$. The second term, 
\begin{align}
G(g)=q  e^{p}  \left[ 
e^{- p e^{-q g} - r g} 
- \frac{e^{q g}}{p^{1+ p}}\,  
\gamma(1 + p,p e^{-q g}) 
\right]\,,
\label{eq:eq5}
\end{align}
arises from gaps formed by spontaneous dissociation. A simplification of the gap-size distribution by invoking the approximation $\gamma(1 + p,p e^{-q g})\approx p^{1+p}e^{-pe^{-qg}-(q+r)g} /(1+p)$ reveals that the distribution crosses over from an exponential form, $G(g)\sim r e^{-rg}$, in the low-collisions limit ($p\to 0$), to a Gaussian form, $G(g)\sim q e^{-\frac{rq}{2} g^2}$ in the high-collisions limit ($p\to \infty$), see Fig.~\ref{fig:overall}(c). 

From the normalization, $\int_0^{\infty}R(g)dg=1$, Eq.~\ref{eq:eq5} implies, $\int_0^{\infty} G(g)dg=1-f_c$, so an increase in the fraction of collisions reduces the fraction of non-zero gaps. To check whether collisions also reduce the typical gap size, we compute the mean gap size, $\langle g \rangle = \int_0^{\infty} g R(g)dg$, which takes the simple form,
\begin{align}
\langle g\rangle = \frac{1}{r}\,.
\label{eq:eq6}
\end{align}
Although non-zero gaps arise due to the spontaneous dissociation, the mean gap size is independent of the spontaneous unbinding rate. However, if we average only over the gaps of non-zero sizes, then the conditional mean gap size is $\langle g \rangle_+ = \int^{\infty}_{0} g G(g)dg / [\int^{\infty}_{0} G(g)dg]= \frac{1}{r(1-f_c)}$, and it can depend on the spontaneous unbinding rate via $f_c$.

\begin{figure}
\centering
\includegraphics[angle=270,scale=0.231]{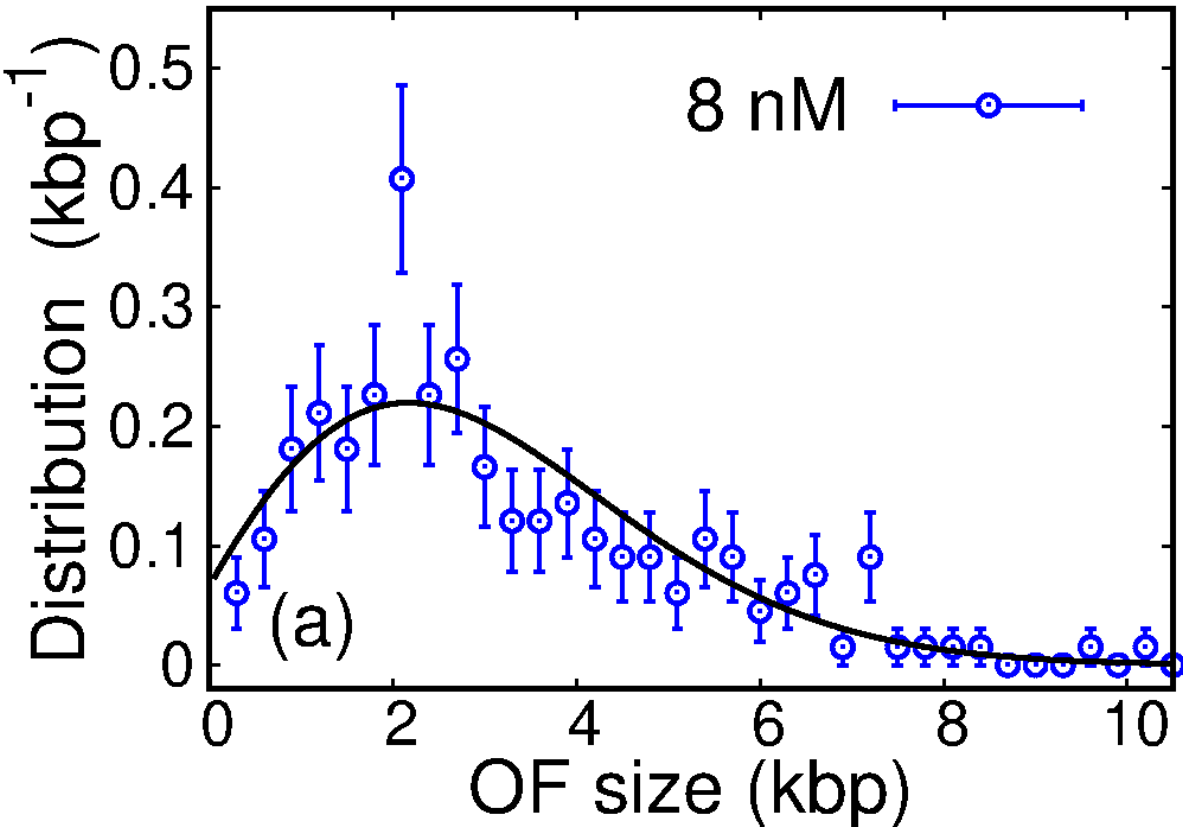}
\includegraphics[angle=270,scale=0.195]{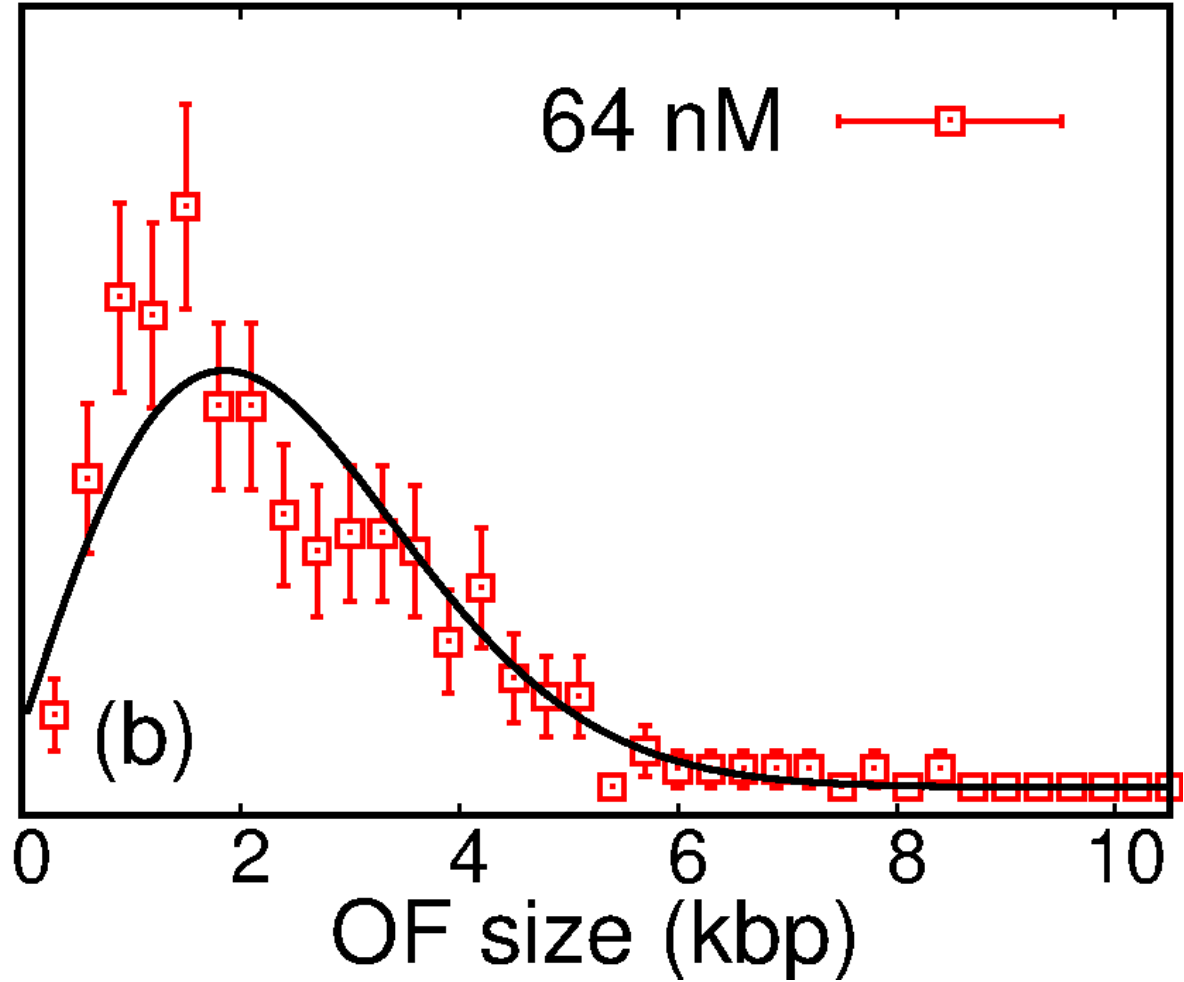}
\caption{Distribution of OF sizes in T4 bacteriophage: 
Symbols represents the experimental data obtained from \cite{chastain2000analysis} for (a) $8\,{\rm nM}$ and  (b) $64\,{\rm nM}$ primase. The error bars represent the standard error; see Appendix \ref{appendix:A} for details. The solid line is from Eq.~\ref{eq:eq1}. }
\label{fig:okazaki}
\end{figure}

We validate our theoretical results (Eqs.~\ref{eq:eq1}--\ref{eq:eq6}) using kinetic Monte Carlo simulations; see Fig.~\ref{fig:overall} for comparison of $f_C$,  $Q(z)$, and $R(g)$ with the simulation results. To evaluate the incomplete gamma functions in Eqs.~\ref{eq:eq2} and \ref{eq:eq5}, we used Mathematica \cite{Mathematica14}.

\section{Comparison with experiments}
To compare our results with experiments, we make use of OFs measured in earlier studies \cite{chastain2000analysis,yang2006control}.  In these experiments, replisomes are reconstituted by assembling the required proteins on circular DNA templates, allowing rolling-circle replication to take place. Thus synthesized OF sizes are determined by radiolabeled nucleotide incorporation followed by gel electrophoresis or electron microscopy. OF have been measured for replisomes of T4 bacteriophage and \textit{B. subtilis} and we now apply the developed framework to study the lagging-strand polymerase dynamics in them.

The T4 bacteriophage replisome is a well-established model system, with experimentally measured OF sizes \cite{chastain2000analysis,chen2013insights,yang2006control}. Ref \cite{chastain2000analysis} reports OF sizes at $8\, {\rm nM}$ and $64\, {\rm nM}$ primase concentrations, measured using electron microscopy and gel electrophoresis. From these data, we compute the empirical OF size distribution and fit Eq.~\ref{eq:eq1} treating $q$ and $r$ as fitting parameters,  see Fig.~\ref{fig:okazaki} for the fits and Appendix \ref{appendix:A} for details of the fitting procedure. From the best fit parameters $(q^*,r^*)$, we estimate the binding rate $\pi^*=vr^*$ and the spontaneous unbinding rate  $\epsilon^*=vq^*$, using a DNA polymerase speed  $400\,{\rm bp \cdot sec^{-1}}$ \cite{chastain2000analysis}. The results summarized in Table~\ref{tab:combined_optimal} show that the binding rate is approximately threefold higher and the spontaneous unbinding rate about twofold smaller at $64\,{\rm nM}$ primase compared to $8\,{\rm nM}$. This indicates that reducing primase concentration primarily affects the binding rate, likely due to slower initiation from reduced priming.

The non-monotonic form of the OF size distribution in Fig.~\ref{fig:okazaki} suggests, as discussed earlier, a possible dominance of collision-governed dissociation in T4 bacteriophage replisomes. Substituting the best-fit parameters into Eq.~\ref{eq:eq2}, we find that as primase concentration increases from $8\,{\rm nM}$ to $64\,{\rm nM}$, the fraction of collisions increases from  79\% to 90\%, consistent with our assertion that polymerase dissociation is largely governed by collisions.

\begin{figure}
\centering
\includegraphics[angle=-90,width=0.30\textwidth]{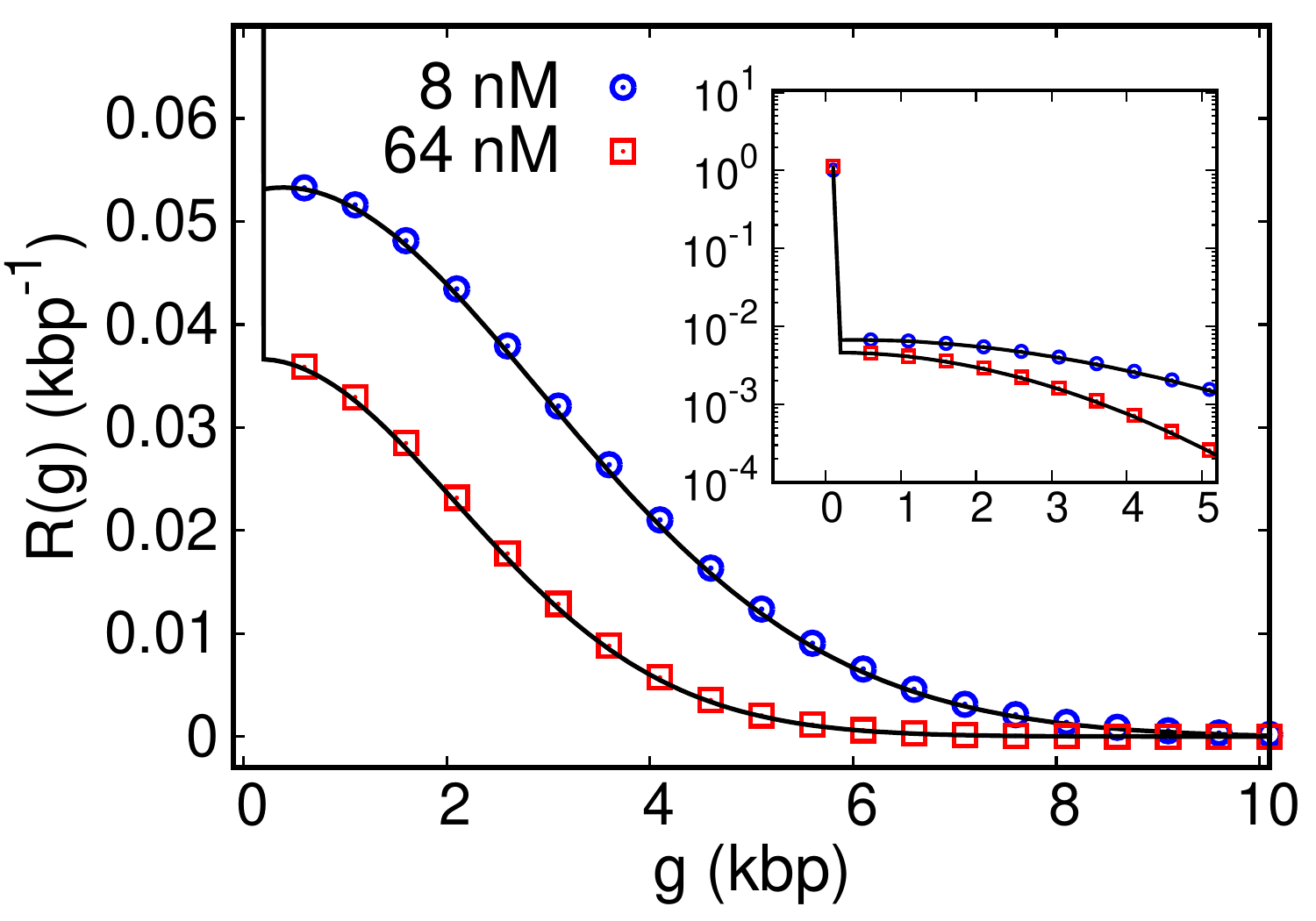}
\caption{
Distribution of gap-size in T4 bacteriophage:  The solid line is from Eq.~\ref{eq:eq4}, and symbols are simulation results for the parameters given in Table~\ref{tab:combined_optimal}.The peak at $g=0$ (shown in the inset) reflects the delta-function (Eq.~\ref{eq:eq4}) contribution from collision-induced dissociations, highlighting that zero-size gaps constitute a significant fraction of the overall gap population compared to non-zero gaps.}
\label{fig:gap_8_64}
\end{figure}

 The gap-size distribution computed from Eq.~\ref{eq:eq4} shifts toward smaller gaps as the primase concentration increases from $8\,{\rm nM}$ to $64\,{\rm nM}$ (see Fig.~\ref{fig:gap_8_64}). The distribution exhibits a kink at $g=0$ (see insets), arising from the delta function in Eq.~\ref{eq:eq4}, which reflects the collision-induced polymerase dissociation. The mean gap size, computed from Eq.~\ref{eq:eq6}, decreases from $476~{\rm bp}$ at $8\,{\rm nM}$ to $164~{\rm bp}$ at $64\,{\rm nM}$. Experimental determination of gap sizes and their distribution is challenging due to formation of secondary structures by the single-stranded DNA, underscoring the significance of this prediction. By hindering the secondary structure formation, gap sizes have been measured for T4 bacteriophage replisomes in \cite{yang2006control}. The observed gap size is approximately 350 bp at 400 nM primase, which is of a similar order of magnitude to the gap size we found.

\begin{figure}
\centering
\includegraphics[width=0.48\textwidth]{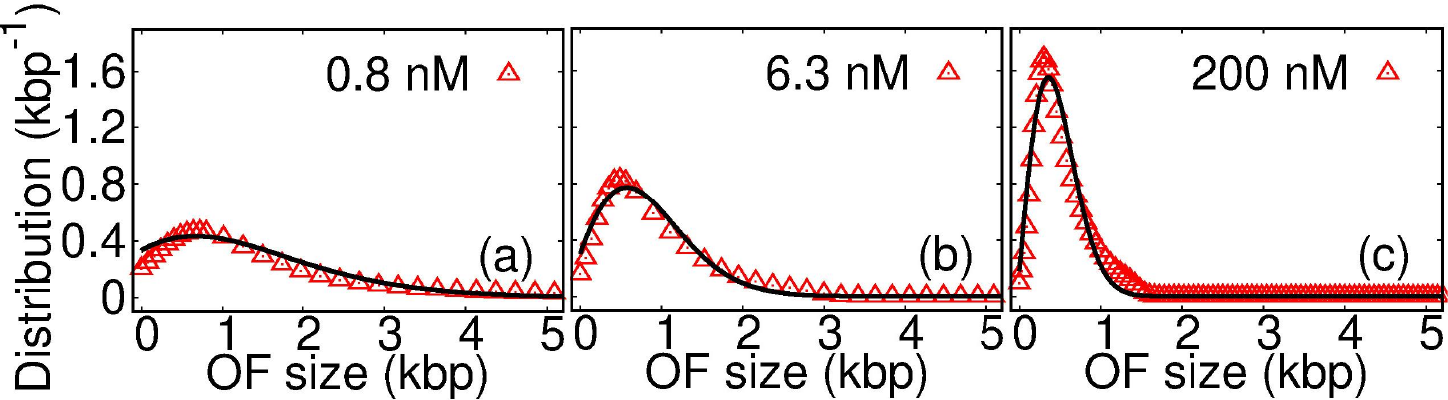}
\caption{
OF size distributions in \textit{B. subtilis:}  The solid line is from Eq.~\ref{eq:eq1}, and symbols are smoothed experimental data extracted from gel electrophoresis images reported in \cite{sanders2010reconstitution}, see Appendix \ref{appendix:b} for details of the data extraction and analysis.}
\label{fig:subtilis-combined}
\end{figure}

\textit{B. subtilis} is another system for which OF sizes have been measured at different primase concentrations using gel electrophoresis \cite{sanders2010reconstitution}. Analysis of these data reveals a gradual increase in the non-monotonicity of the OF size distribution with increasing primase concentration, suggesting a higher fraction of collisions at elevated primase levels (see Fig.~\ref{fig:subtilis-combined} and Appendix \ref{appendix:b} for details). From the best fit parameters, we confirm that the fraction of collisions increases from 0.28 at $0.4~{\rm nM}$ to 0.92 at $200~{\rm nM}$ (Table.~\ref{tab:combined_optimal}).

\begin{table}
\caption{ \label{tab:combined_optimal}
Model parameters for T4 bacteriophage and \textit{B. subtilis} at varying primase concentrations. We find $f_c$, $\langle z \rangle$, and $\langle g \rangle$ from Eqs\,.\ref{eq:eq2}, \ref{eq:eq3}, and  \ref{eq:eq6}, respectively.}
\begin{ruledtabular}
\begin{tabular}{ccccccc}
Organism & Primase & $\epsilon^*$ & $\pi^*$ & $\langle z \rangle$ & $\langle g \rangle$ & $f_c$ \\
 & (nM) & ($\mathrm{sec^{-1}}$) & ($\mathrm{sec^{-1}}$) & (bp) & (bp) & \\
\hline

\multirow{2}{*}{T4}
& 8  & 0.026 & 0.84 & 3058 & 476 & 0.79 \\
& 64 & 0.016 & 2.4  & 2415 & 164 & 0.90 \\

\hline

\multirow{10}{*}{\centering \textit{B. subtilis}}
& 0.4  & 0.166 & 0.17   & 2404 & 3333 & 0.29 \\
& 0.8  & 0.186 & 0.73   & 1440 & 769 & 0.52\\
& 1.6  & 0.176 & 1.68   & 1083 & 333  & 0.67 \\
& 3.1  & 0.173 & 2.52   & 922 & 222  & 0.72 \\
& 6.3  & 0.167 & 3.19  & 846 & 175 & 0.75 \\
& 12.5  & 0.171 & 3.59  & 747  & 156  & 0.76 \\
& 25    & 0.175 & 4.47  & 712  & 125 & 0.78 \\
& 50    & 0.162 & 6.38   & 640  & 88   & 0.82 \\
& 100   & 0.160  & 9.07   & 544  & 62   & 0.85 \\
& 200   & 0.096 & 22.23  & 468  & 25   & 0.92 \\

\end{tabular}
\end{ruledtabular}
\vspace{1mm}
{\footnotesize Polymerase speed: T7, $400\,\mathrm{bp\,s^{-1}}$ \cite{chastain2000analysis}; 
\textit{B. subtilis}, $560\,\mathrm{bp\,s^{-1}}$ \cite{sanders2010reconstitution}.}
\end{table}

\section{Outline of Calculations}
We now outline the computation of $Q(z)$ and $R(g)$ in Eqs.~\ref{eq:eq1} and \ref{eq:eq4}, respectively. The key quantity essential for this computation is the transition probability $W(z,g|z',g')$, defined as the probability that the current OF and gap sizes are $z$ and $g$, given that the preceding OF and gap sizes were $z'$ and $g'$. It satisfies the normalization, $\iint^{\infty}_{0} dz\, dg\, W(z,g|z',g')=1$. A detailed derivation of it is given in the Appendix \ref{appendix:c}; its explicit form is,
\begin{align}
W(z,g|z',g')=\theta(g+z-z') rq e^{-(q+r)z -r(g-z')} \nonumber \\
+ \theta(z-z')\delta(g)r e^{-r(z-z')-qz} \,,
\label{eq:eq7}
\end{align}
where the step function $\theta(x)$ is defined such that it is 1 when $x\geq 0$ and 0 otherwise. The first term corresponds to spontaneous polymerase dissociation, yielding non-zero gaps, while the second term arises from collision-induced dissociation with preceding OFs, resulting in zero gaps.

We now define an n-point conditional probability, $P_{n}(z,g|0)$, which is the probability for the polymerase to synthesize n'th OF of size $z$ and leave a gap of size $g$ for given initial conditions. The initial conditions are set by the bound or unbound state of the polymerase at the replication origin ($n=0$). However, this dependence is not essential, as we focus on the steady state in which the system loses memory of the initial conditions. 

We express n-point conditional probability in terms of the transitional probability, $W(z,g|z',g')$,  via the two-dimensional Chapman-Kolmogorov equation \cite{risken1989fokker}, $P_{n}(z,g|0) = \iint^\infty_0 P_{n-1}(z',g'|0) W(z,g|z',g') dz'dg'$. When $q>0$, $P_n(z,g|0)$ is independent of $n$ in the large $n$ limit. Therefore, we define the steady state probability, $P(z,g) \equiv \lim_{n \to \infty} P_{n}(z,g|0)$, which satisfies
\begin{align}
     P(z,g) = \iint^\infty_0 P(z',g') W(z,g|z',g') dz'dg'\,. \label{eq:eq8}
\end{align}

Marginalizing $P(z,g)$ over $g$ yields the OF size distribution, $Q(z)= \int^\infty_0 P(z,g) dg$. Integrating Eq.~\ref{eq:eq8} with respect $g$ and using that $W(z,g|z',g')$ is independent of $g'$, we arrive at the following one-dimensional Chapman-Kolmogorov equation for $Q(z)$,
\begin{align}
     Q(z) = \int^z_0 Q(z') w_f(z|z') dz' + \int^{\infty}_z Q(z') w_b(z|z') dz'\, \label{eq:eq9}
\end{align}
where $w_b(z|z') = q \, e^{-q z} $ and $w_f(z|z')=  (q+r) \, e^{-q z - r (z-z')} $. 

Equation~\ref{eq:eq9} is analogous to the Chapman-Kolmogorov equation of a continuous-space random walker with position-dependent hopping probabilities \cite{van1992stochastic}. We solve Eq.~\ref{eq:eq9}  using the Laplace transform technique \cite{ARFKEN2013963}; the solution is given in Eq.~\ref{eq:eq1}, see Appendix \ref{appendix:d} for the details of the derivation. 

Similarly, marginalizing $P(z,g)$ over $g$ gives the gap-size distribution, $R(g)=\int_0^{\infty}P(z,g)dz$. Integrating Eq.~\ref{eq:eq9} with respect to $z$, $R(g)$ can be expressed in terms of $Q(z)$ as 
\begin{align}
     R(g) = \int^\infty_0 Q(z') w(g|z') dz'\,, \label{eq:eq10}
\end{align}
where $w(g|z')= \frac{r}{r+q} e^{-qz'} \delta(g)+ \frac{qr}{q+r}[\theta(z'-g)e^{-q(z'-g)}+\theta(g-z')e^{-r(g-z')}]$. Substituting the OF size distribution from  Eq.~\ref{eq:eq1} into Eq.~\ref{eq:eq10} and carrying out the integration, we obtain the gap-size distribution in Eq.~\ref{eq:eq4}. In Eq.~\ref{eq:eq4},  the coefficient of the delta function corresponds to the fraction of collisions given in Eq.~\ref{eq:eq2}. 
\section{Conclusions}

In summary, we developed a biophysical model for semi-discontinuous DNA replication by incorporating stochastic dynamics of the lagging-strand polymerase. We computed the size distribution of OFs and the intervening gaps. Our central result is that collision-induced polymerase dissociation primarily determines the shape of these distributions. In particular, the distribution of OF sizes becomes non-monotonic at higher collision fractions, a feature that enables the identification of the dominant dissociation mechanism from the experimentally measured OFs.

By applying the model to the T4 bacteriophage replisome\cite{chen2013insights} and \emph{B. subtilis}~\cite{sanders2010reconstitution}, we found that an increase in primase concentration increases the binding rate, thereby increasing the fraction of collision-induced dissociation and leading to stronger non-monotonic behavior. The increased binding rate with the primase concentration also results in a decrease in the mean OF and gap sizes.  A decrease in the OF size \cite{wu1992coordinated,yang2006control,seco2013,graham2017independent} and number of non-zero gaps \cite{graham2017independent} with increasing primase concentration has been observed in other systems as well. Our results suggest that these trends are driven by an increased fraction of collision-induced dissociation. 

Finally, the present model has several limitations. For instance, experimental studies have indicated the possibility of the speed asymmetry between leading- and lagging-strand polymerases  \cite{georgescu2014replisome} and as well as mechanical coupling between them \cite{graham2017independent,kurth2013solution}. Polymerase exchange between the replisome and the cytoplasm has been observed in some cases \cite{Beattie2017}. Incorporating these effects into the present framework and exploring their consequences will be an interesting direction for future work.

\begin{acknowledgments}
We thank Manoj Gopalakrishnan for the insightful discussions, Simone Pigolotti for critical comments on the manuscript, and  VIT Vellore for providing the computational facility. DB gratefully acknowledges support from the VIT Seed Grant (SG20220060). JG acknowledges the Inspire fellowship from the Department of
Science and Technology (DST), Government of India (No. DST/INSPIRE Fellowship/2023/IF230119).
\end{acknowledgments}

\bibliographystyle{apsrev4-2}
\bibliography{references}

@article{graham2017independent,
  title={Independent and stochastic action of DNA polymerases in the replisome},
  author={Graham, James E and Marians, Kenneth J and Kowalczykowski, Stephen C},
  journal={Cell},
  volume={169},
  number={7},
  pages={1201--1213},
  year={2017},
  publisher={Elsevier},
  doi = {https://doi.org/10.1016/j.cell.2017.05.041}
}

@book{van1992stochastic,
title = {Stochastic Processes in Physics and Chemistry},
publisher = {Elsevier},
address = {Amsterdam},
year = {2007},
series = {North-Holland Personal Library},
issn = {09255818},
doi = {https://doi.org/10.1016/B978-044452965-7/50007-6},
url = {https://www.sciencedirect.com/science/article/pii/B9780444529657500076},
author = {N.G. {Van Kampen}}
}

@article{mueser2010structural,
  title={Structural analysis of bacteriophage T4 DNA replication: a review in the Virology Journal series on bacteriophage T4 and its relatives},
  author={Mueser, Timothy C and Hinerman, Jennifer M and Devos, Juliette M and Boyer, Ryan A and Williams, Kandace J},
  journal={Virol. J.},
  volume={7},
  number={1},
  pages={359},
  year={2010},
doi = {https://doi.org/10.1186/1743-422X-7-359},
  publisher={Springer}
}

@book{risken1989fokker,
    author="Risken, Hannes",
    title="The Fokker-Planck Equation: Methods of Solution and Applications",
    year="1996",
    publisher="Springer",
    address="Berlin",
    isbn="978-3-642-61544-3",
    doi="10.1007/978-3-642-61544-3_2",
    url="https://doi.org/10.1007/978-3-642-61544-3_2"
}

@article{sun2023okazaki,
  title={Okazaki fragment maturation: DNA flap dynamics for cell proliferation and survival},
  author={Sun, Haitao and Ma, Lingzi and Tsai, Ya-Fang and Abeywardana, Tharindu and Shen, Binghui and Zheng, Li},
  journal={Trends Cell Biol.},
  volume={33},
  number={3},
  pages={221--234},
  year={2023},
  publisher={Elsevier},
doi = {10.1016/j.tcb.2022.06.014}
}

@article{williams2021high,
  title={High-fidelity DNA ligation enforces accurate Okazaki fragment maturation during DNA replication},
  author={Williams, Jessica S and Tumbale, Percy P and Arana, Mercedes E and Rana, Julian A and Williams, R Scott and Kunkel, Thomas A},
  journal={Nat. Commun.},
  volume={12},
  number={1},
  pages={482},
  year={2021},
  publisher={Nature Publishing Group UK London},
  doi = {https://doi.org/10.1038/s41467-020-20800-1}
}

@article{alberts1987prokaryotic,
author = {Alberts, Bruce Michael },
title = {Prokaryotic DNA replication mechanisms},
journal = {Philos. Trans. R. Soc. Lond. B Biol. Sci.s},
volume = {317},
number = {1187},
pages = {395-420},
year = {1987},
doi = {10.1098/rstb.1987.0068},
URL = {https://royalsocietypublishing.org/doi/abs/10.1098/rstb.1987.0068},
}

@article{benkovic2001replisome,
  title={Replisome-mediated DNA replication},
  author={Benkovic, Stephen J and Valentine, Ann M and Salinas, Frank},
  journal={Annu. Rev. Biochem.},
  volume={70},
  number={1},
  pages={181--208},
  year={2001},
  publisher={Annual Reviews 4139 El Camino Way, PO Box 10139, Palo Alto, CA 94303-0139, USA},
   doi = {https://doi.org/10.1146/annurev.biochem.70.1.181},
}

@article{balakrishnan2010reconstitution,
  title={Reconstitution of eukaryotic lagging strand DNA replication},
  author={Balakrishnan, Lata and Gloor, Jason W and Bambara, Robert A},
  journal={Methods},
  volume={51},
  number={3},
  pages={347--357},
  year={2010},
  publisher={Elsevier},
  doi = {10.1016/j.ymeth.2010.02.017}
}

@article{kurth2013solution,
  title={A solution to release twisted DNA during chromosome replication by coupled DNA polymerases},
  author={Kurth, Isabel and Georgescu, Roxana E and O'Donnell, Mike E},
  journal={Nature},
  volume={496},
  number={7443},
  pages={119--122},
  year={2013},
  doi={https://doi.org/10.1038/nature11988},
  publisher={Nature Publishing Group UK London}
}

@article{okazaki1968mechanism,
author = {R Okazaki  and T Okazaki  and K Sakabe  and K Sugimoto  and A Sugino },
title = {Mechanism of DNA chain growth. I. Possible discontinuity and unusual secondary structure of newly synthesized chains.},
journal = {Proc. Natl. Acad. Sci.},
volume = {59},
number = {2},
pages = {598-605},
year = {1968},
doi = {10.1073/pnas.59.2.598},
URL = {https://www.pnas.org/doi/abs/10.1073/pnas.59.2.598},
}

@article{chastain2000analysis,
  title={Analysis of the Okazaki fragment distributions along single long DNAs replicated by the bacteriophage T4 proteins},
  author={Chastain, Paul D and Makhov, Alexander M and Nossal, Nancy G and Griffith, Jack D},
  journal={Mol. Cell},
  volume={6},
  number={4},
  pages={803--814},
  year={2000},
  publisher={Elsevier},
  doi = {10.1016/S1097-2765(05)00093-6}
}

@article{wu1992coordinated,
  title={Coordinated leading-and lagging-strand synthesis at the Escherichia coli DNA replication fork. V. Primase action regulates the cycle of Okazaki fragment synthesis.},
  author={Wu, CA and Zechner, EL and Reems, JA and McHenry, CS and Marians, KJ},
  journal={J. Biol. Chem.},
  doi ={https://doi.org/10.1016/S0021-9258(19)50632-9},
  volume={267},
  number={6},
  pages={4074--4083},
  year={1992},
  publisher={Elsevier}
}

@article{hacker1994rapid,
  title={The rapid dissociation of the T4 DNA polymerase holoenzyme when stopped by a DNA hairpin helix. A model for polymerase release following the termination of each Okazaki fragment.},
  author={Hacker, Kevin J and Alberts, Bruce M},
  journal={J. Biol. Chem.},
  volume={269},
  number={39},
  pages={24221--24228},
  year={1994},
    doi = {https://doi.org/10.1016/S0021-9258(19)51071-7},
    url = {https://www.sciencedirect.com/science/article/pii/S0021925819510717},
  publisher={Elsevier}
}

@article{yang2006control,
  title={The control mechanism for lagging strand polymerase recycling during bacteriophage T4 DNA replication},
  author={Yang, Jingsong and Nelson, Scott W and Benkovic, Stephen J},
  journal={Mol. Cell},
  volume={21},
  number={2},
  pages={153--164},
  year={2006},
doi = {https://doi.org/10.1016/j.molcel.2005.11.029},
url = {https://www.sciencedirect.com/science/article/pii/S1097276505018447},
  publisher={Elsevier}
}

@article{chen2013insights,
  title={Insights into Okazaki fragment synthesis by the T4 replisome: the fate of lagging-strand holoenzyme components and their influence on Okazaki fragment size},
  author={Chen, Danqi and Yue, Hongjun and Spiering, Michelle M and Benkovic, Stephen J},
  journal={J. Biol. Chem.},
  volume={288},
  number={29},
  pages={20807--20816},
  year={2013},
  publisher={Elsevier},
doi = {10.1074/jbc.M113.485961}
}

@article{spiering2017rna,
  title={RNA primer--primase complexes serve as the signal for polymerase recycling and Okazaki fragment initiation in T4 phage DNA replication},
  author={Spiering, Michelle M and Hanoian, Philip and Gannavaram, Swathi and Benkovic, Stephen J},
  journal={Proc. Natl. Acad. Sci.},
  volume={114},
  number={22},
  pages={5635--5640},
  year={2017},
  publisher={National Academy of Sciences},
doi = {https://doi.org/10.1073/pnas.1620459114}
}

@article{nelson2008rna,
  title={RNA primer handoff in bacteriophage T4 DNA replication: the role of single-stranded DNA-binding protein and polymerase accessory proteins},
  author={Nelson, Scott W and Kumar, Ravindra and Benkovic, Stephen J},
  journal={J. Biol. Chem.},
  volume={283},
  number={33},
  pages={22838--22846},
  year={2008},
  publisher={Elsevier},
  doi = {https://doi.org/10.1074/jbc.M802762200}
}

@article{
Katarzyna2018,
author = {Katarzyna H. Maslowska  and Karolina Makiela-Dzbenska  and Jin-Yao Mo  and Iwona J. Fijalkowska  and Roel M. Schaaper },
title = {High-accuracy lagging-strand DNA replication mediated by DNA polymerase dissociation},
journal = {Proc. Natl. Acad. Sci.},
volume = {115},
number = {16},
pages = {4212-4217},
year = {2018},
doi = {10.1073/pnas.1720353115},
URL = {https://www.pnas.org/doi/abs/10.1073/pnas.1720353115},
}

@article {Beattie2017,
article_type = {journal},
title = {Frequent exchange of the DNA polymerase during bacterial chromosome replication},
author = {Beattie, Thomas R and Kapadia, Nitin and Nicolas, Emilien and Uphoff, Stephan and Wollman, Adam JM and Leake, Mark C and Reyes-Lamothe, Rodrigo},
editor = {Berger, James M},
volume = 6,
year = 2017,
month = {mar},
pub_date = {2017-03-31},
pages = {e21763},
citation = {eLife 2017;6:e21763},
doi = {10.7554/eLife.21763},
url = {https://doi.org/10.7554/eLife.21763},
journal = {eLife},
issn = {2050-084X},
publisher = {eLife Sciences Publications, Ltd},
}

@book{ARFKEN2013963,
editor = {George B. Arfken and Hans J. Weber and Frank E. Harris},
title = {Mathematical Methods for Physicists (Seventh Edition)},
publisher = {Academic Press},
address = {Boston},
year = {2013},
isbn = {978-0-12-384654-9},
doi = {https://doi.org/10.1016/B978-0-12-384654-9.00020-7},
url = {https://www.sciencedirect.com/science/article/pii/B9780123846549000207},
author = {George B. Arfken and Hans J. Weber and Frank E. Harris}
}

@book{Wasserman2004,
author="Wasserman, Larry",
title="All of Statistics: A Concise Course in Statistical Inference",
year="2004",
publisher="Springer",
address="New York",
isbn="978-0-387-21736-9",
doi="10.1007/978-0-387-21736-9_9",
url="https://doi.org/10.1007/978-0-387-21736-9_9"
}

@book{Alberts2014,
  title={Molecular Biology of the Cell},
  author={Alberts, B. and Heald, R. and Johnson, A. and Morgan, D. and Raff, M. and Roberts, K. and Walter, P.},
  isbn={9781315735368},
  doi = {https://doi.org/10.1201/9781315735368},
  year={2022},
  publisher={W.W. Norton and Company}
}

@article{Burgers2017,
   author = "Burgers, Peter M.J. and Kunkel, Thomas A.",
   title = "Eukaryotic DNA Replication Fork", 
   journal= "Annu. Rev. Biochem.",
   year = "2017",
   volume = "86",
   number = "Volume 86, 2017",
   pages = "417-438",
   doi = "https://doi.org/10.1146/annurev-biochem-061516-044709",
   url = "https://www.annualreviews.org/content/journals/10.1146/annurev-biochem-061516-044709",
   publisher = "Annual Reviews",
   issn = "1545-4509",
   type = "Journal Article",
  }

@article{Johnson2005,
   author = "Johnson, Aaron and O'Donnell, Mike",
   title = "CELLULAR DNA REPLICASES: Components and Dynamics at the Replication Fork", 
   journal= "Annu. Rev. Biochem.",
   year = "2005",
   volume = "74",
   number = "Volume 74, 2005",
   pages = "283-315",
   doi = "https://doi.org/10.1146/annurev.biochem.73.011303.073859",
   url = "https://www.annualreviews.org/content/journals/10.1146/annurev.biochem.73.011303.073859",
   publisher = "Annual Reviews",
   issn = "1545-4509",
   type = "Journal Article",
}

@article{Yao2009,
author = {Nina Y. Yao  and Roxana E. Georgescu  and Jeff Finkelstein  and Michael E. O'Donnell },
title = {Single-molecule analysis reveals that the lagging strand increases replisome processivity but slows replication fork progression},
journal = {Proc. Natl. Acad. Sci.},
volume = {106},
number = {32},
pages = {13236-13241},
year = {2009},
doi = {10.1073/pnas.0906157106},
URL = {https://www.pnas.org/doi/abs/10.1073/pnas.0906157106},
}

@article{Yuan2013,
    author = {Yuan, Quan and McHenry, Charles S.},
    title = {Cycling of the E. coli lagging strand polymerase is triggered exclusively by the availability of a new primer at the replication fork},
    journal = {Nucleic Acids Res.},
    volume = {42},
    number = {3},
    pages = {1747-1756},
    year = {2013},
    month = {11},
    issn = {0305-1048},
    doi = {10.1093/nar/gkt1098},
    url = {https://doi.org/10.1093/nar/gkt1098},
}

@article{Kelly2019,
author = {Thomas Kelly  and A. John Callegari },
title = {Dynamics of DNA replication in a eukaryotic cell},
journal = {Proc. Natl. Acad. Sci.},
volume = {116},
number = {11},
pages = {4973-4982},
year = {2019},
doi = {10.1073/pnas.1818680116},
URL = {https://www.pnas.org/doi/abs/10.1073/pnas.1818680116},
}

@article{Sahoo2021,
  title = {Accuracy and speed of elongation in a minimal model of DNA replication},
  author = {Sahoo, M. and N., Arsha and Baral, P. R. and Klumpp, S.},
  journal = {Phys. Rev. E},
  volume = {104},
  issue = {3},
  pages = {034417},
  numpages = {8},
  year = {2021},
  month = {Sep},
  publisher = {American Physical Society},
  doi = {10.1103/PhysRevE.104.034417},
  url = {https://link.aps.org/doi/10.1103/PhysRevE.104.034417}
}

@article {Bhat2022,
article_type = {journal},
title = {Speed variations of bacterial replisomes},
author = {Bhat, Deepak and Hauf, Samuel and Plessy, Charles and Yokobayashi, Yohei and Pigolotti, Simone},
editor = {Nourmohammad, Armita and Tyler, Jessica K and Amir, Ariel},
volume = 11,
year = 2022,
month = {jul},
pub_date = {2022-07-25},
pages = {e75884},
citation = {eLife 2022;11:e75884},
doi = {10.7554/eLife.75884},
url = {https://doi.org/10.7554/eLife.75884},
journal = {eLife},
issn = {2050-084X},
publisher = {eLife Sciences Publications, Ltd},
}

@article{Karschau2012,
title = "Optimal Placement of Origins for DNA Replication",
author = "Jens Karschau and Blow, J. Julian and de Moura, Alessandro P. S.",
year = "2012",
month = jan,
day = "30",
doi = "10.1103/PhysRevLett.108.058101",
language = "English",
volume = "108",
pages = "--",
journal = "Phys. Rev. Lett.",
issn = "0031-9007",
publisher = "American Physical Society",
number = "5",
}

@article{Gauthier2010,
  title = {Defects and DNA Replication},
  author = {Gauthier, Michel G. and Herrick, John and Bechhoefer, John},
  journal = {Phys. Rev. Lett.},
  volume = {104},
  issue = {21},
  pages = {218104},
  numpages = {4},
  year = {2010},
  month = {May},
  publisher = {American Physical Society},
  doi = {10.1103/PhysRevLett.104.218104},
  url = {https://link.aps.org/doi/10.1103/PhysRevLett.104.218104}
}

@article{Sharma2013,
  title = {Error correction during DNA replication},
  author = {Sharma, Ajeet K. and Chowdhury, Debashish},
  journal = {Phys. Rev. E},
  volume = {86},
  issue = {1},
  pages = {011913},
  numpages = {7},
  year = {2012},
  month = {Jul},
  publisher = {American Physical Society},
  doi = {10.1103/PhysRevE.86.011913},
  url = {https://link.aps.org/doi/10.1103/PhysRevE.86.011913}
}

@article{Retkute2012,
  title = {Mathematical modeling of genome replication},
  author = {Retkute, Renata and Nieduszynski, Conrad A. and de Moura, Alessandro},
  journal = {Phys. Rev. E},
  volume = {86},
  issue = {3},
  pages = {031916},
  numpages = {15},
  year = {2012},
  month = {Sep},
  publisher = {American Physical Society},
  doi = {10.1103/PhysRevE.86.031916},
}

@article{Retkute2011,
  title = {Dynamics of DNA Replication in Yeast},
  author = {Retkute, Renata and Nieduszynski, Conrad A. and de Moura, Alessandro},
  journal = {Phys. Rev. Lett.},
  volume = {107},
  issue = {6},
  pages = {068103},
  numpages = {4},
  year = {2011},
  month = {Aug},
  publisher = {American Physical Society},
  doi = {10.1103/PhysRevLett.107.068103},
  url = {https://link.aps.org/doi/10.1103/PhysRevLett.107.068103}
}

@article{reijns2015,
  title={Lagging-strand replication shapes the mutational landscape of the genome},
  author={Reijns, Martin AM and Kemp, Harriet and Ding, James and Marion de Proc{\'e}, Sophie and Jackson, Andrew P and Taylor, Martin S},
  journal={Nature},
  volume={518},
  number={7540},
  pages={502--506},
  year={2015},
  publisher={Nature Publishing Group UK London},
  doi= {https://doi.org/10.1038/nature14183}
}

@article{zheng2011,
  title={Fen1 mutations that specifically disrupt its interaction with PCNA cause aneuploidy-associated cancer},
  author={Zheng, Li and Dai, Huifang and Hegde, Muralidhar L and Zhou, Mian and Guo, Zhigang and Wu, Xiwei and Wu, Jun and Su, Lei and Zhong, Xueyan and Mitra, Sankar and Huang, Qin and Kernstine, Kemp H  and Pfeifer, Gerd P and Shen, Binghui},
  journal={Cell Res.},
  doi = {https://doi.org/10.1038/cr.2011.35},
  volume={21},
  number={7},
  pages={1052--1067},
  year={2011},
  publisher={Nature Publishing Group}
}

@article{seco2013,
  title={Bacteriophage SPP1 DNA replication strategies promote viral and disable host replication in vitro},
  author={Seco, Elena M and Zinder, John C and Manhart, Carol M and Lo Piano, Ambra and McHenry, Charles S and Ayora, Silvia},
  journal={Nucleic Acids Res.},
  volume={41},
  number={3},
  pages={1711--1721},
  year={2013},
  publisher={Oxford University Press},
  doi={10.1093/nar/gks1290}
}

@article{georgescu2014replisome,
  title={Replisome mechanics: lagging strand events that influence speed and processivity},
  author={Georgescu, Roxana E and Yao, Nina and Indiani, Chiara and Yurieva, Olga and O'Donnell, Mike E},
  journal={Nucleic Acids Res.},
  doi = {https://doi.org/10.1093/nar/gku257},
  volume={42},
  number={10},
  pages={6497--6510},
  year={2014},
  publisher={Oxford University Press}
}

@article{duderstadt2014replication,
  title={Replication-fork dynamics},
  author={Duderstadt, Karl E and Reyes-Lamothe, Rodrigo and van Oijen, Antoine M and Sherratt, David J},
  journal={Cold Spring Harb. Perspect. Biol.},
  volume={6},
  number={1},
  doi = {https://doi.org/10.1101/cshperspect.a010157},
  pages={a010157},
  year={2014},
  publisher={Cold Spring Harbor Lab}
}

@article{Moura2024,
  title = {Mathematical model for the distribution of DNA replication origins},
  author = {de Moura, Alessandro and Karschau, Jens},
  journal = {Phys. Rev. E},
  volume = {110},
  issue = {3},
  pages = {034408},
  numpages = {12},
  year = {2024},
  month = {Sep},
  publisher = {American Physical Society},
  doi = {10.1103/PhysRevE.110.034408},
  url = {https://link.aps.org/doi/10.1103/PhysRevE.110.034408}
}

@misc{Mathematica14,
  author = {Wolfram Research{,} Inc.},
  title = {Mathematica, {V}ersion 14.3},
  url = {https://www.wolfram.com/mathematica},
  note = {Champaign, IL, 2025}
}

@article{sanders2010reconstitution,
  title={Reconstitution of the B. subtilis replisome with 13 proteins including two distinct replicases},
  author={Sanders, Glenn M and Dallmann, H Garry and McHenry, Charles S},
  journal = {Mol. Cell},
  volume={37},
  number={2},
  pages={273--281},
  year={2010},
  doi={https://doi.org/10.1016/j.molcel.2009.12.025},
  publisher={Elsevier}
}

@article{Schneider_Rasband_Eliceiri_2012, 
title={NIH image to imagej: 25 years of image analysis}, 
url={https://www.nature.com/articles/nmeth.2089}, 
journal = {Nat. Methods}, 
volume = {9},
pages = {671–675},
publisher={Nature Publishing Group}, 
author={Schneider, Caroline A and Rasband, Wayne S and Eliceiri, Kevin W}, 
year={2012}
}

@article{tanner2008single,
  title={Single-molecule studies of fork dynamics in Escherichia coli DNA replication},
  author={Tanner, Nathan A and Hamdan, Samir M and Jergic, Slobodan and Loscha, Karin V and Schaeffer, Patrick M and Dixon, Nicholas E and Van Oijen, Antoine M},
 journal = {Nat. Struct. Mol. Biol.},
  volume={15},
  number={2},
  pages={170--176},
  year={2008},
  doi={https://doi.org/10.1038/nsmb0908-998a},
  publisher={Nature Publishing Group US New York}
}

@article{Fay1982,
title = {Size classes of products synthesized processively by two subassemblies of Escherichia coli DNA polymerase III holoenzyme.},
 journal={J. Biol. Chem.},volume = {257},
number = {10},
pages = {5692-5699},
year = {1982},
issn = {0021-9258},
doi = {https://doi.org/10.1016/S0021-9258(19)83833-4},
url = {https://www.sciencedirect.com/science/article/pii/S0021925819838334},
author = {P J Fay and K O Johanson and C S McHenry and R A Bambara},
}

\appendix
\section{Analysis of OF sizes of T4 bacteriophage}
\label{appendix:A}
We analyzed the OF size data of T4 bacteriophage replication system reported in an earlier experimental study \cite{chastain2000analysis}. In that work, individual fragment lengths were measured by electron microscopy at primase concentrations of $8\,{\rm nM}$ and $64\,{\rm nM}$. The number of observed OFs  is $N=221$ at $8\,{\rm nM}$ and $N=261$ at $64\,{\rm nM}$.  The smallest dected OF is $0.1 \,{\rm kbp}$, while the largest fragment size $(z_{\rm max})$ is $10.6\,{\rm kbp}$ and $8.2\,{\rm kbp}$ for  $8\,{\rm nM}$  and $64\,{\rm nM}$, respectively. 

We bin the OF sizes separately for each primase concentration at every bin size $dz=300\,{\rm bp}$. The maximum number of bins, $I= 1+\lfloor z_{\rm max}/dz \rfloor$, is $36$ for $8\,{\rm nM}$ and $28$ for $64\,{\rm nM}$. From these data, we construct an empirical size distribution of OF sizes, $P_{\rm exp}(z)$, which is defined as the fraction of OFs between $z$ and $z+dz$ per unit length. If  $n(z)$ is the number of OFs of sizes between $z$ and $z+dz$, then   $P_{\rm exp}(z)=n(z)/(N dz)$. Treating $n(z)$ as a Poisson random variable, we compute the associated standard error  $\sigma(z)= \sqrt{n}/(N dz)$.

To determine the optimal parameters for which $Q(z)$ best fits $P_{\rm exp}(z)$, we employ the Maximum-likelihood estimation method \cite{Wasserman2004}. To that end, we define a Gaussian likelihood function
\begin{align}
\mathcal{L} = \prod^{I}_{i=1} \frac{1}{\sqrt{2 \pi \sigma(i dz)}} e^{-\frac{\left[P_{\rm exp}(i dz )-Q(i dz)\right]^2}{2 \sigma(z)^2}}\,,
\end{align}
where, $Q(z)$ is the OF size distribution given in Eq.~\ref{eq:eq1}. We find the optimal parameters $(q^*, r^*)$ by maximizing the log-likelihood function $\log \mathcal{L}$;  see Table~\ref{tab:optimal-t4} for the parameter values. 

\begin{table}[h!]
\caption{\label{tab:optimal-t4}
Best fit parameters ($q^*, r^*$) for T4 bacteriophage.}
\begin{ruledtabular}
\begin{tabular}{c|cc}
Primase (nM) & $q^*$ ($\times 10^{-5}~\mathrm{bp^{-1}}$)  & $r^*$ ($\times 10^{-3}~\mathrm{bp^{-1}}$)\\ \hline 
8 & $6.6$ & $2.1$\\

64 & $4.05$ & $6.09$ \\
\end{tabular}
\end{ruledtabular}
\end{table}

\section{Analysis of OF sizes of \textit{B. subtilis}}
\label{appendix:b}

\begin{figure*}
\centering
\includegraphics[width=2.0\columnwidth]{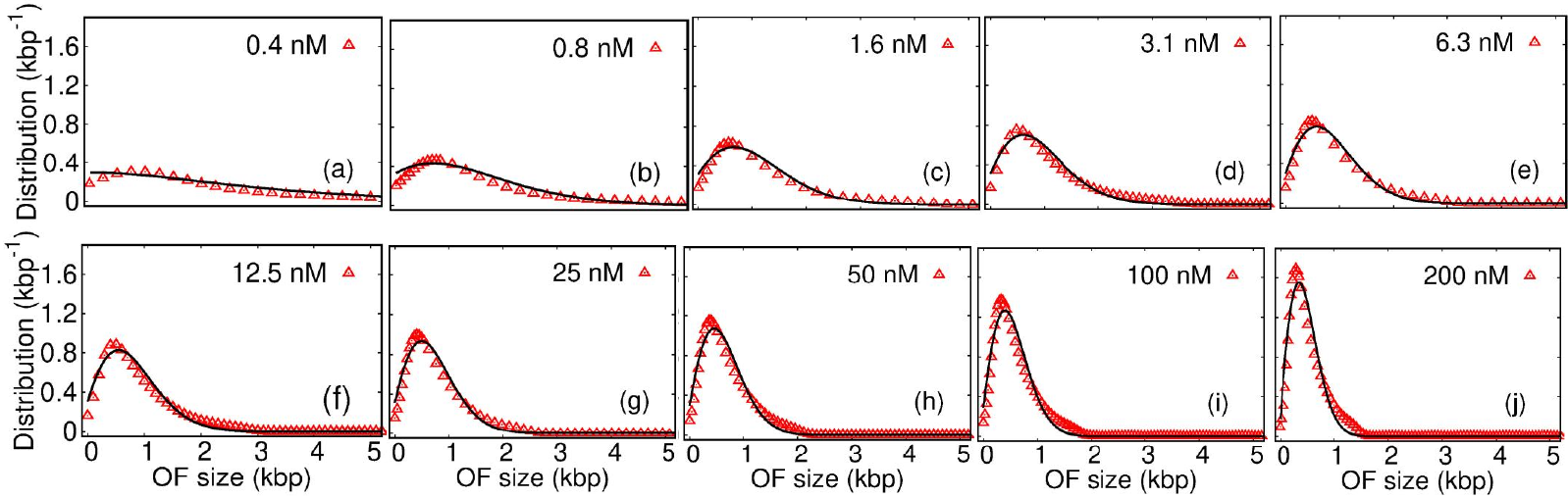}
   \caption{OF size distributions for \textit{B. subtilis} under varying primase concentrations are shown, as indicated in each panel. The solid line is from Eq.~\ref{eq:eq1}, while symbols denote the empirical OF size distributions obtained from gel electrophoresis images reported in \cite{sanders2010reconstitution}. Figure~\ref{fig:subtilis-combined} shows three representative images from this set, corresponding to $0.8~{\rm nM}$, $6.3~{\rm nM}$, and $200~{\rm nM}$.
   }
\label{figure: all primase concentration} 
\end{figure*}

In \textit{B. subtilis}, individual OF sizes from electron microscopy are not directly available; consequently, we cannot apply the maximum-likelihood estimation used for T4 bacteriophage. However, \cite{sanders2010reconstitution} reports OF sizes quantified using gel electrophoresis. In that study, OF sizes are measured across twofold dilutions of primase concentration, ranging from $200,\mathrm{nM}$ to $0.4,\mathrm{nM}$. We extract intensity profiles from the gel electrophoresis images using \textit{ImageJ} \cite{Schneider_Rasband_Eliceiri_2012}. The resulting profiles exhibit a flat tail region, which likely arises from background fluorescence \cite{Schneider_Rasband_Eliceiri_2012}. To avoid overfitting, we exclude intensity values below 20\% of the peak from the analysis. We then obtain the empirical OF size distribution from the processed intensity profiles using kernel density estimation \cite{Mathematica14}.  Finally, we determine the optimal parameters $q^*$ and $r^*$ by minimizing the least-squares difference between $Q(z)$ given in Eq.~\ref{eq:eq1} and the empirical OF size distribution. We present the empirical OF size distributions along with the corresponding fits in Fig.~\ref{figure: all primase concentration}. The corresponding optimal parameter values are listed in Table~\ref{tab:optimal-bsubtilis}.

As in the case of T4 bacteriophage, the binding rate increases with increasing primase concentration, while the unbinding rate remains nearly constant (see Table~\ref{tab:optimal-bsubtilis}). This results in an increase in the fraction of collisions with the primase concentration as shown in Table~\ref{tab:combined_optimal}.

\begin{table}[h]
\caption{\label{tab:optimal-bsubtilis} Best fit parameters ($q^*, r^*$) for \textit{B. subtilis}}
\begin{ruledtabular}
\begin{tabular}{c|cc}
Primase (nM) & $q^*$ ($\times 10^{-4}~\mathrm{bp^{-1}}$) & $r^*$ ($\times 10^{-3} ~\mathrm{bp^{-1}}$) \\
\hline
0.4   & $2.97$ & $0.3$ \\
0.8   & $3.32$ & $1.3$ \\
1.6   & $3.14$ & $3.0$ \\
3.1   & $3.08$ & $4.5$\\
6.3   & $2.99$  & $5.7$\\
12.5  & $3.05$  & $6.4$  \\
25    & $3.12$  & $8.0$  \\
50    & $2.89$ & $11.4$  \\
100   & $2.85$ & $16.2$  \\
200   & $1.71$ &  $39.7$ \\
\end{tabular}
\end{ruledtabular}
\vspace{1mm}
\end{table}

\section{Derivation of the transition probability} 
\label{appendix:c}
To derive the transition probability, $W(z,g \mid z',g')$, we analyze the dynamics of a lagging-strand polymerase within our model. Consider a polymerase that unbinds after synthesizing an OF of size $z'$ and leaving a gap $g'$. Let $\tau$ be the waiting time period for the polymerase to rebind again near the replisome on the lagging strand and initiate synthesis of the next OF. In that time period, the replisome progresses a distance $v \tau$ along the template. If $z$ denotes the size of the next OF and $g$ the corresponding gap left by the polymerase, then they must satisfy the relation:
\begin{align}
z+g=z'+ v \tau\,.
\label{eq:eq12}
\end{align}
In our model, we assumed that binding occurs with rate $\pi$. Therefore, the waiting time, $\tau$, is a random variable with the distribution 
\begin{align}
\phi(\tau) = \theta(\tau)\pi e^{-\pi \tau}
\end{align}
 where $\theta(\tau)$ is a step function defined in the main text. 
 
 In addition, if the polymerase does not collide with the preceding OF, spontaneous unbinding of the polymerase occurs with a rate $\epsilon$. The size distribution of $z$, under spontaneous unbinding is $\theta(z)qe^{-qz}$, with $q=\epsilon/v$. Spontaneous unbinding corresponds to the scenario $g>0$; therefore, from Eq.~\ref{eq:eq12}  $z<z'+v\tau$. In contrast, collision-induced dissociation corresponds to $g=0$; therefore, from Eq.~\ref{eq:eq12}  $z=z'+v\tau$. 
 
 Combining these, the size distribution of OF for given $(z',g', \tau)$ is $qe^{-qz}$ for $z<z'+ v \tau$, and  $ \delta(z-z' - v \tau)e^{-q(z'+v \tau)}$ otherwise. Using the condition in Eq.~\ref{eq:eq12}, the joint distribution of OF size, $z$, and gap size, $g$, for given $(z',g', \tau)$ is  $qe^{-qz}\delta(z+g-z'-v\tau)\theta(z'+ v \tau-z)+ \delta(z-z' - v \tau)\delta(g) e^{-q(z'+v \tau)}$. By averaging it with respect to $\phi(\tau)$, we obtain the transition probability in Eq.~\ref{eq:eq8}

\section{Computation of  OF size distribution using the Laplace Transform method} 
\label{appendix:d}
To compute the steady-state distribution of the OF sizes, $Q(z)$, we employ the Laplace transform technique. We define the Laplace transform of any function $f(z)$, for $z>0$, as follows, $\tilde{f}(s)=\int^{\infty}_{0}e^{-sz}f(z)dz$. Computing the Laplace transform of $Q(z)$ in Eq.~\ref{eq:eq9}, we obtain the following recursive relation,
\begin{equation}
    \tilde{Q} (s) = B(s)~\tilde{Q} (s + q)  + C(s)
\end{equation}
where, $B(s) = [(q+r)/(s+q+r)] - [q/(s+q)]$, and $C(s) = q/(s+q) $. We define a raising operator, $\hat{E}$, such that $\hat{E}\tilde{f}(s)=\tilde{f}(s+q)$. The solution can be written using the raising operator as $\tilde{Q}(s) =  \sum_{n=0}^\infty  \left[B(s) \hat{E}\right]^n C(s)$, and its explicit form is
\begin{align}
   \tilde{Q}(s) = \sum_{n=0}^\infty  \frac{ q r^n s  (s+r)  }{(s+nq)[s+(n+1)q]\prod_{m=0}^n (s+mq+r)}\,.
\end{align}
Inverting the Laplace transform using standard techniques \cite{ARFKEN2013963}, we obtain the OF size distribution in the form of a series,
\begin{equation*}
Q(z) =  \sum_{j=1}^{\infty} \sum_{n=0}^{\infty}  \frac{ (j+p)q p^{k+n}  e^{-(jq+r)z} }{ (p-n) (n+1-p) \prod_{m=1,j\not =m}^{n+j} (m-j)}\,.
\end{equation*}
Simplification by carrying out the summation yields,
\begin{equation}
Q(z) = \left[q+r(1- e^{-qz} )\right] e^{-(q+r)z + p (1 - e^{-qz}) }\,,
\end{equation}
which is the same as the OF size distribution in Eq.~\ref{eq:eq1}.

\end{document}